\documentclass[twocolumn,tightenlines,prc,aps,showpacs,showkeys,groupedaddress,floatfix]{revtex4}
\usepackage{epsfig}%
\usepackage{graphicx}
\usepackage{dcolumn}
\usepackage{bm}
\usepackage{longtable}%
\begin{document}
\title{$\alpha$-$\alpha$ Double Folding Cluster Potential Description of the $^{12}$C+$^{24}$Mg System}
\author{M. Karakoc and I. Boztosun}
\affiliation{Department of Physics, Erciyes University, Kayseri, Turkey}
\date{\today}
\begin{abstract}
We present a simultaneous analysis of the elastic scattering and
fusion cross-section data of the $^{12}$C+$^{24}$Mg system around
the Coulomb barrier and over energies by using the microscopic
$\alpha$-$\alpha$ double folding cluster potential within the
framework of the optical model and the coupled-channels formalism.
The $\alpha$-$\alpha$ double folding cluster potential is obtained
by using the $\alpha$-cluster distribution densities of the nuclei
in the usual double folding procedure. The microscopic potential
results are compared with the findings of the phenomenological deep
and shallow potentials. It is subsequently shown that only
phenomenological deep real, microscopic nucleon-nucleon and
$\alpha$-$\alpha$ double folding cluster potentials provide a
consistent description of the angular distributions and fusion
cross-section data simultaneously. The effect of the inclusion of
the excited states of the target nucleus $^{24}$Mg on the fusion
cross-section predictions is also determined by the coupled-channels
calculations, which are shown to improve the agreement.
\end{abstract}
 \maketitle
\section{Introduction}
Determining the shape of the nuclear potential between two colliding
pairs is a long-standing problem. The theoretical investigations of
the precisely measured experimental data at high energies well over
the Coulomb barrier for systems like $^{12}$C+$^{12}$C and
$^{16}$O+$^{16}$O have led to the determination of the gross
features of the local Optical potentials. Subsequently, ambiguities
have been clarified in many cases regarding the depths of the real
parts of the nuclear potentials \cite{Sat97}. However, it is not yet
possible to claim the same conclusive arguments for the shape of the
nuclear potential for the reactions around the Coulomb barrier. The
theoretical analysis suffers from a number of serious drawbacks such
as the failure to determine the shape of the interaction potential,
the reproduction of the oscillatory structure and the out-of-phase
problem between theoretical predictions and experimental data.

In this context, the $^{12}$C+$^{24}$Mg reaction
\cite{Sci97,lep93,Fil89,Car76,Bozcmg} has been extensively
investigated both experimentally and theoretically. The conventional
optical model analysis conducted so far  fails to explain all or
some of the experimental data by using shallow or deep optical
potentials \cite{Sci97,lep93,Fil89}. Moreover, there has been no
detailed microscopic study using folding models attempting to
explain the individual angular distributions and fusion
cross-sections data simultaneously. Therefore, we aim to analyze the
$^{12}$C+$^{24}$Mg system for energies from 16.0 to 24.0 MeV by
using the $\alpha$-$\alpha$ microscopic double folding cluster (DFC)
potential. Our results are shown in comparison with the
nucleon-nucleon double folding (NN-DF), phenomenological shallow
(WS$_S$) and deep (WS$_{D}^{2}$) real potentials.

In the next section, we introduce the potentials used in the optical
model and coupled-channels (CC) formalism. In sections \ref{results}
and \ref{ccmodel}, the optical and CC results are shown. Section
\ref{conc} is devoted to our conclusion.
\section{The Optical Model}
\label{model} In order to make a comparative study of this reaction,
we have used four different potentials for the real part of the
optical model potential: Two are microscopic, which are calculated
from microscopic NN-DF and $\alpha$-$\alpha$ DFC potentials and the
other two are phenomenological deep and shallow potentials. We
provide the details of the $\alpha$-$\alpha$ DFC potential and leave
the NN-DF and phenomenological potentials to references provided in
\cite{Sci97,Boziop,mesuttez}. The projectile and target nuclei,
which we study in this paper consist of integer multiple of the
alpha particles. It has been known that 4n type nuclei have an
$\alpha$-cluster structure \cite{rae,freer}. Therefore, it will be
very interesting to obtain the interaction potential by considering
the alpha-particle structure of these nuclei.  For this purpose, the
$\alpha$-$\alpha$ DFC potential is constructed in a similar way to
the ordinary DF one: We fold an $\alpha$-$\alpha$ effective
interaction with $\alpha$-clusters distribution densities and
formulate the nucleus-nucleus DFC optical model potential
\cite{Azab2} as
\begin {equation}
V_{DFC}(r)=\int\int{\rho_{cP}(r_{1})\rho_{cT}(r_{2})\nu_{\alpha\alpha}
(|\vec{r}+\vec{r}_{2}-\vec{r}_{1}|)d^{3}r_{1}d^{3}r_{2}} \label
{VDFC}
\end {equation}
where $\rho_{cP}$ and $\rho_{cT}$ are the $\alpha$-cluster
distributions for projectile and target nuclei and
$\nu_{\alpha\alpha}$ is the effective $\alpha$-$\alpha$ interaction.

The matter distribution of a nucleus is known and can be obtained
from:
\begin {equation}
\rho_{M}(r)=\rho_{0M}(1+wr^{2})exp(-\beta r^{2}) \label {rhoM}
\end {equation}
This is a modified form of the Gaussian shape for the $\rho_{M}$(r),
projectile and target densities. The matter density of an $\alpha$
nucleus can also be obtained from:
\begin {equation}
\rho_{\alpha}(r)=\rho_{0\alpha}exp(-\beta r^{2}) \label {rhoalpha}
\end {equation}
The parameters for the $\rho_{0\alpha}$, $\rho_{0M}$, $w$ and
$\beta$ used in equations \ref{rhoM} and \ref{rhoalpha} are given in
Table \ref{density}.
\begin{table}\caption{The parameters of nuclear matter
densities of the $^{12}$C and $^{4}$He \cite{Azab1}. The parameter
of the $^{24}$Mg nuclear matter density is obtained from RIPL-2
\cite{ripl2}.} \label{density}
\begin{ruledtabular}
\begin{tabular}{cccccc}
 Nuclei &$\rho_{0}$&w&$\beta$& $<r^{2}>^{1/2}$\\
      &(fm$^{-3}$) &(fm$^{-2}$)&(fm$^{-2}$) & (fm)\\\hline
 $^{12}$C  &0.1644&0.4988&0.3741&2.407\\
 $^{24}$Mg &0.1499&0.4012&0.2383&3.050\\
 $^{4}$He  &0.4229&   0  &0.7024&1.460
\end{tabular}
\end{ruledtabular}
\end{table}
If $\rho_{c}(r')$ is the $\alpha$-cluster distributions function
inside the nucleus, then we can relate the nuclear matter density
distribution functions of the nucleus, $\rho_{M}$(r), to that of the
$\alpha$-particle nucleus, $\rho_{\alpha}$(r), as
\begin {equation}
\rho_{M}(r)=\int{\rho_{c}(r')\rho_{\alpha}(|\vec{r}-\vec{r'}|)d^{3}r'}
\label {get_cluster}
\end {equation}
Since the densities of the nucleus and the alpha particle can be
calculated from  equations \ref{rhoM} and \ref{rhoalpha}, by using
Fourier transform techniques \cite{Sat79} for expression (\ref
{get_cluster}), we can obtain the $\alpha$-cluster distribution
function $\rho_{c}(r')$ as:
\begin {equation}
\rho_{c}(r')=\rho_{0c}(1+\mu r'^{2})exp(-\xi r'^{ 2})
\label{rhocluster}
\end {equation}
with $\eta = \lambda - \beta \label {eta_const}, \quad \xi = \beta
\lambda / \eta \label {xi_const}, \quad
\mu=\frac{2w\lambda^{2}}{\eta(2\eta-3w)}$. 
Inserting this $\alpha$-cluster distribution together with the
following effective $\alpha$-$\alpha$ interaction potential of Buck
{\it et al.} \cite{Buck}, we can obtain the $\alpha$-$\alpha$ DFC
from equation \ref{VDFC}.
\begin {equation}
\nu_{\alpha\alpha}(r)=-122.6225exp(-0.22r^{2})
\label {Vaa}
\end {equation}

For the phenomenological potentials, we use slightly modified
versions of the potentials previously conducted for this reaction.
For the deep potential, we use the potentials of Boztosun and Rae
\cite{Bozcmg} and for the shallow potential, we use the potentials
of Sciani \emph{et al} \cite{Sci97}.

The parameters of the potentials are given in Table~\ref{pot}. The
codes Dfpot~\cite{dfpot} and Fresco \cite{fresco} are used for all
calculations.
\section{Results and Discussions}
\label{results}

The experimental data of the $^{12}$C+$^{24}$Mg reaction has been
analyzed in the laboratory system from 16.0 to 24.0 MeV by using
both phenomenological and microscopic potentials within the
above-described optical model.

In order to obtain the best fit between the experimental data and
the theoretical calculations, we have conducted a $\chi^{2}$ search
to define the parameters of the potentials. For the microscopic DF
potentials, we have two free parameters: N$_{R}$ and W$_{0}$. The
normalization factor (N$_{R}$) of the real part  and the depth
(W$_{0}$) of the imaginary part have been varied on a grid and the
results of this systematic search have shown that the N$_{R}$ or
W$_{0}$ parameters cannot be varied continuously and still produce
equally satisfying fits. For the normalization factor of
$\alpha$-$\alpha$ DFC potential, the lowest $\chi^{2}$ values are
generally obtained between 0.7 and 0.9, but we have chosen the
parameter N$_{R_{\alpha-\alpha}}$=0.72, which provides a consistent
description for all energies. For the NN-DF potential,
N$_{R_{NN}}$=0.84.

\begin{figure}
\epsfxsize 8.5cm \centerline{\epsfbox{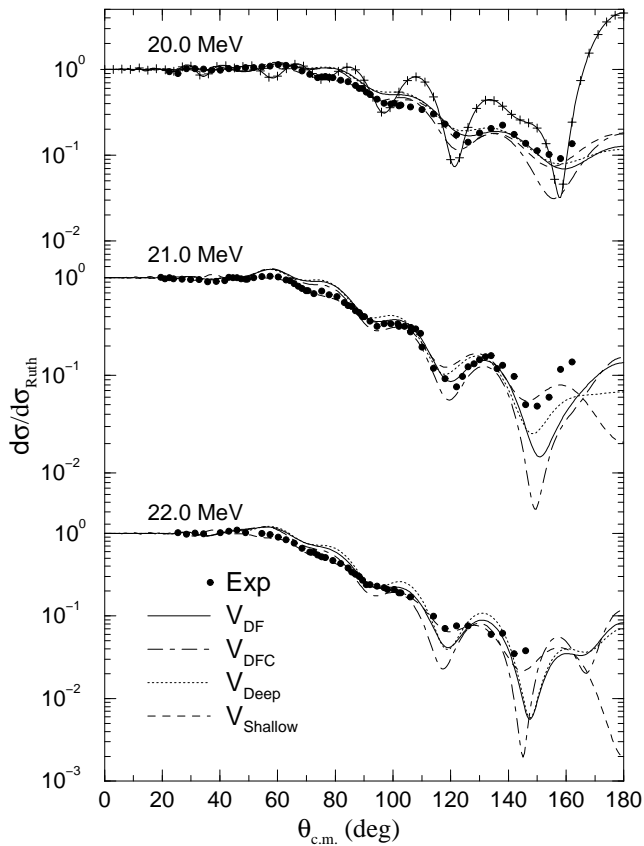}} \vskip-0.35cm
\caption{OM elastic scattering results obtained by using NN-DF
(solid lines), $\alpha$-$\alpha$ DFC (dot dashed lines),
phenomenological deep (dotted lines) and shallow (dashed lines)
potentials. The solid line with plus shows the CC prediction of the
shallow potential using the short range imaginary potential at
E$_{Lab}$=20MeV (see the text). Experimental data is from
\cite{Sci97}.} \label{angular1}
\end{figure}

\begin{table}
\caption{The parameters of the real and imaginary potentials. All
imaginary potentials have WS volume shape.} \label{pot}
\begin{ruledtabular}
\begin{tabular}{ccccccc}
            Pot.&$V_{0}$&$r_{V}$&$a_{V}$&$W_{0}$&$r_{W}$&$a_{W}$\\
            Type       & (MeV)   & (fm)   & (fm)     & (MeV)              & (fm)    & (fm)   \\\hline
           NN       &  -               &  -   &   -    &$1.8E+1.6$ &0.30  &  0.286 \\
$\alpha$-$\alpha$  &  -               &  -   &   -    &$3.7E-43.4$&0.30  &  0.286 \\
  $WS^{2}_{D}$      &427.0             &0.88 & 1.187  &$0.4E+30.0$&0.30  &  0.286 \\
  $WS_{S}$      &$49.1-0.56E$&1.29 & 0.400  & $0.054E-0.47$ &1.77  &  0.600
\end{tabular}
\end{ruledtabular}
\end{table}
Some of the results of our analysis obtained by using microscopic
and phenomenological potentials are shown in Figure \ref{angular1}
for the individual angular distributions and in Figure
\ref{xsecReaction} for the fusion cross-section data. Numerical
values at energies where the experimental data are available, are
also shown in Table \ref{fus-xsec} for the fusion cross-section.

We may infer from the figures that the theoretical results obtained
by using the microscopic and phenomenological potentials present
more or less the same behavior: It is difficult to see the
difference at forward angles since they overlap. The difference
becomes apparent at large angles. However, the lowest $\chi^{2}$
values for the individual angular distributions are provided by the
shallow real potential. If we perceive the lowest $\chi^{2}$ values
as the best fit, then we may say that the shallow potential provides
the best fit. If we look at the figures, we also perceive that the
theoretical results obtained by using the shallow potential give
very good agreement with the experimental data at forward,
intermediate and large angles. The magnitude of the cross-section is
correctly provided and the minima/maxima are at the correct places
with phases.

However, the same shallow potential that explains the angular
distribution is unable to predict the fusion cross-section. The
theoretical calculation for the fusion cross-section is almost twice
as big as the experimental data.

\begin{table}
\caption{Theoretical reaction and experimental fusion
cross-sections. Experimental data are taken from
\cite{Iwamato,Gary}.} \label{fus-xsec}
\begin{ruledtabular}
\begin{tabular}{cccccc}
 &&\quad $\sigma (mb)$\\\hline
 E (MeV) &Exp.&NN-DF&$\alpha$-$\alpha$ DFC & WS$_{D}^{2}$ &WS$_{S}$ \\\hline
 20.0 &198.82 &243.23  &273.66  &236.11  &432.50  \\
 21.0 &243.56 &320.64  &354.45  &311.83  &530.18  \\
 22.0 &331.73 &393.56  &430.47  &371.41  &678.13  \\
 23.0 &426.43 &456.69  &493.40  &443.54  &723.56  \\
 24.0 &435.01 &518.43  &553.70  &493.81  &791.99
\end{tabular}
\end{ruledtabular}
\end{table}
Nevertheless, the deep potentials, both microscopic and
phenomenological ones, provide a good agreement for the individual
angular distributions with acceptable $\chi^{2}$ values and predict
the fusion cross-section reasonably well. The magnitude of the
cross-section is correctly predicted and the minima/maxima are at
the correct places with phases. The only discrepancy is at large
angles where minima are predicted deeper than the measured data.
\section{Coupled-Channels Calculations} \label{ccmodel}
The optical model calculations in the previous section provide the
total reaction cross-section, but not the fusion, therefore, there
is a discrepancy between theoretical predictions and the
experimental fusion cross-section data. In order to obtain the
fusion cross-section and improve the agreement, we have to either
use a model-independent approach such as the one used by references
\cite{holdeman1,yamaya} or remove the non-elastic cross-section from
reaction cross-section calculations; we have used the CC model for
this purpose and in our calculations, the fusion cross-section is
obtained in the following way:
\begin{equation}
\sigma_F=\sigma_R-\sigma_{in} \label{fus}
\end{equation}
where $\sigma_F$ denotes the fusion, $\sigma_R$ is the total
reaction and $\sigma_{in}$, the non-elastic cross-sections. In the
present CC calculations, we describe the interaction between
$^{12}$C and $^{24}$Mg nuclei with a deformed optical potential. The
real potential is assumed to have the square of a Woods-Saxon and
the imaginary potential has the standard Woods-Saxon volume shape
\cite{Bozcmg}.

We assume that the target nucleus $^{24}$Mg has a static quadrupole
deformation, and that its rotation can be described within the
framework of the rotational model by deforming the real potential in
the following way
\begin{equation}
R(\theta,\phi)=r_{0}A_{p}^{1/3}+r_{0}A_{t}^{1/3}[1+\beta_{2}
Y_{20}(\theta,\phi)]
\end{equation}
where $p$ and $t$ refer to projectile and target nuclei respectively
and $\beta_{2}$ is the deformation parameter of $^{24}$Mg. We shall
use the exact value of $\beta_{2}$, derived from the deformation
length $\delta$=1.50 fm ($\beta$=0.52). For the Coulomb deformation,
we assume $\beta_{2}^{C}$=$\beta_{2}^{N}$.
\begin{figure}
\epsfxsize 8.5cm \centerline{\epsfbox{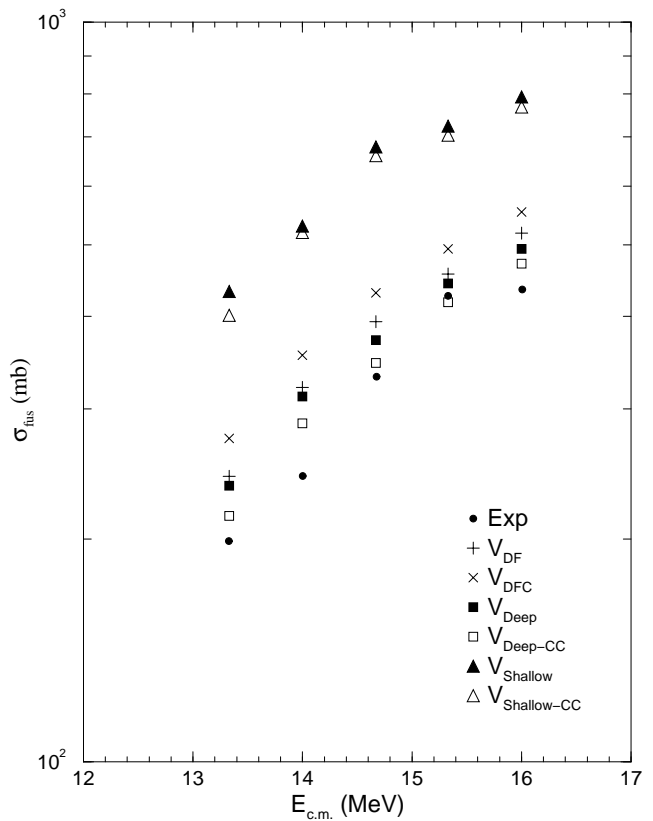}} \vskip-0.4cm
\caption{Experimental fusion cross-sections (filled circles)
\cite{Iwamato,Gary} are compared with theoretically calculated
results using the NN-DF (plus), the $\alpha$-$\alpha$ DFC (cross),
phenomenological deep (filled squares) and shallow (filled
triangle-up) potentials. The coupled channels calculation results
using the phenomenological deep (squares) and shallow (triangle-up)
potentials are also shown. }\label{xsecReaction}
\end{figure}

The results of the CC calculations are shown in
Figure~\ref{xsecReaction} and in Table~\ref{fus-xsec2} for the
fusion calculations. We do not show the elastic scattering angular
distribution results obtained since we are mainly interested in the
effect of the CC calculation on the fusion data prediction.
Nevertheless, the CC results for the elastic scattering angular
distribution are reasonably well. From Figure \ref{xsecReaction}, it
is clear that the inclusion of the 2$^{+}$ and 4$^{+}$ excited
states of $^{24}$Mg affects the calculations and gives a better
agreement for the fusion calculations in comparison with the optical
model. The numerical values of this effect can be seen from Table
\ref{fus-xsec2}. In this table, we present the optical and CC
results for the deep Woods-Saxon squared phenomenological and
shallow Woods-Saxon volume potentials. The inclusion of the 2$^{+}$
and 4$^{+}$ excited states of $^{24}$Mg removes flux from the
elastic channel and the CC results for the fusion data are in better
agreement than the optical model one. From Table \ref{fus-xsec2}, it
may be observed that, while the optical model prediction is around
$\sigma_{F}$=236.11mb at E$_{Lab}$=20.0 MeV, it becomes
$\sigma_{F}$=215.08mb after the inclusion of the 2$^{+}$ and 4$^{+}$
excited states of $^{24}$Mg. The effect is around 22.85mb, which
makes the theoretical CC prediction better in agreement with the
experimental data. The same effect has been also observed for the
shallow real potential, the coupled-channel calculations improve the
agreement with the experimental fusion data, but it is still far
from being comparable with the prediction of the deep potentials.

\begin{table}
\caption{Comparison of the CC and optical model (OM) fusion
cross-section predictions using deep (D) and shallow (S) potentials.
Experimental data are from \cite{Iwamato,Gary}.}
 \label{fus-xsec2}
\begin{ruledtabular}
\begin{tabular}{cccccccc} 
\multicolumn{8}{p{220pt}}{\hspace{4cm}$\sigma(mb)$}  \\
\hline E &Exp.   & OM$_{D}$ & \multicolumn{2}{p{50pt}}{\qquad CC$_{D}$} & OM$_{S}$ & \multicolumn{2}{p{50pt}}{\qquad CC$_{S}$}  \\
\hline
  (MeV)  &       & $\sigma_{F}$ &$\sigma_{F}$&$\sigma_{2^{+}}$& $\sigma_{F}$&$\sigma_{F}$&$\sigma_{2^{+}}$ \\\hline
   20.0  &198.82 &236.11      &215.08       &22.85           &432.50 &401.49       &26.80            \\
   21.0  &243.56 &311.83      &286.72       &27.10           &530.18 &519.42       &31.64            \\
   22.0  &331.73 &371.41      &346.04       &29.86           &678.13 &659.74       &31.71            \\
   23.0  &426.43 &443.54      &418.00       &32.33           &723.56 &703.22       &39.64            \\
   24.0  &435.01 &493.81      &471.67       &34.10           &791.99 &767.74       &45.05            \\
\end{tabular}
\end{ruledtabular}
\end{table}
We have noticed that the failure of the shallow potential may be
related to the long-range imaginary potential we have used in the
calculations. Because of this long range imaginary potential, we
cannot obtain satisfactory agreement with the fusion cross-section.
However, when we reduce the range of the imaginary potential and use
the one similar to the deep potential model, we get a better
agreement with the fusion data, but this time we are unable to
obtain a good agreement with the elastic scattering angular
distribution. This is illustrated in Figure~\ref{angular1} at
E$_{Lab}$=20.0 MeV. In this figure, the solid line with plus shows
the prediction of a short ranged imaginary potential as used in the
deep Woods-Saxon squared potential. The parameters are given in
Table \ref{pot}. As a result, we have reached the conclusion that it
is not possible to explain the elastic scattering angular
distribution and fusion cross-section simultaneously by using a
shallow real potential in contrast with the deep one.
\section{Summary and Conclusion}
\label{conc}
The theoretical description of the $^{12}$C+$^{24}$Mg system has
been very difficult since the experimental data show very
oscillatory features near the Coulomb barrier at very low energies
and a striking backward rise and oscillatory features at forward,
intermediate and backward angles at high energies. In this paper, we
have shown a consistent description of the elastic scattering of the
$^{12}$C+$^{24}$Mg system at energies around the Coulomb barrier and
over, from 16.0 MeV to 24.0 MeV, in the laboratory system by using
the NN-DF and $\alpha$-$\alpha$ DFC potentials in the Optical model
calculations. This constitutes the first detailed application of the
folding model. All potentials, both deep and shallow, have provided
excellent agreement with the experimental data for the elastic
scattering individual angular distributions at different laboratory
energies; however, only deep potentials explain the angular
distributions and fusion cross-section data simultaneously.  As we
have argued in the paper, the origin of the large difference between
deep and shallow potentials for the fusion cross-section data is
related to the long-range imaginary potential. This work clearly
demonstrates the inadequacy of using shallow potentials in
describing such nuclear reactions and underlines the validity of the
double folding potentials.

This project is supported by T\"{U}B\.{I}TAK, Grant No: TBAG-2398
and Erciyes University, FBT-04-15.

\end{document}